\documentclass[12pt]{article}
\usepackage{amsmath,amssymb,amstext,amsthm,exscale,latexsym}
\input epsf
\newcommand {\al}   {\alpha}       \newcommand {\bt}  {\beta}
\newcommand {\g }   {\gamma}       \newcommand {\G }  {\Gamma}
\newcommand {\dl}   {\delta}       \newcommand {\e }  {\epsilon}
        
\newcommand {\ve}   {\varepsilon}  
\newcommand {\lm}   {\lambda}

\newcommand {\vf }  {\varphi}      
         \newcommand {\om}  {\omega}

\newcommand {\pl}   {\partial}     \newcommand {\nb}  {\nabla}
\newcommand   {\const}{{\sf\,const}}

\theoremstyle{definition}

\begin{document}
\title     {String model with dynamical geometry  \\
            and torsion}
\author    {M. O. Katanaev
            \thanks{E-mail: katanaev@mi.ras.ru}
            \ and I. V. Volovich
            \thanks{E-mail: volovich@mi.ras.ru}\\ \\
            \sl Steklov Mathematical Institute,\\
            \sl Gubkin St. 8, 119991, Moscow}
\date {25 April 1986}
\maketitle
\begin{abstract}
A string model with dynamical metric and torsion is proposed. The
geometry of the string is described by an effective Lagrangian for the
scalar and vector fields. The path integral quantization of the string
is considered.
\end{abstract}

1. Superstring theory now appears as a candidate for a unified theory
of elementary particles [1--5].
\nocite{GreSch84,CaHoStWi85,GrHaMaRo85,Weinbe85,AreVol85}%
The string concept still plays an important role in strong interaction
and statistical physics [6--10].
\nocite{BrDiHo76,Polyak81A,FraTse82,AmDuFr85,Olesen85}%
The fundamental problem of the bosonic string model is that it may be
quantized only in 26 spacetime dimensions. Furthermore it contains a
tachyon state. In this note we consider a new bosonic string model which
has some appealing properties and perhaps circumvents the difficulties
mentioned.

The classical string model is defined by the Lagrangian \cite{BrDiHo76}
\begin{equation}                                        \label{estrla}
  L_0=\frac12\sqrt{g}g^{\al\bt}\pl_\al X^\mu\pl_\bt X_\mu,~~~~
  g=\det g_{\al\bt},
\end{equation}
which contains the string position variable $X^\mu(\zeta)$, $\mu=1,\dotsc,D$,
and the metric tensor $g_{\al\bt}(\zeta)$ on the two-dimensional space of
the parameters $\zeta^\al$, $\al=0,1$, which describe the string. The metric
$g_{\al\bt}$ acts like a Lagrange multiplier in (\ref{estrla}) and may be
eliminated algebraically via its equation of motion.

The bosonic part of the action for known string models is essentially the
same and is proportional to the area of the string surface. This is true
for (\ref{estrla}) after elimination of the metric via its equation of
motion. The action for (\ref{estrla}) is invariant under (i) reparametrization
of the string surface or general coordinate transformation of the parameters
$\zeta^\al$, (ii) Poincar\'e transformation of the string position variable
$X^\mu$, (iii) and Weyl transformation of the metric $g_{\al\bt}$.

We try to extend the list of possible string models from the
geometric point of view. It seems that there is no way to
generalize the Lagrangian (\ref{estrla}) without introducing the
internal degrees of freedom for a string. One commonly
acknowledged generalization is fermionic or superstring theory
which attributes the string with fermionic degrees of freedom in a
supersymmetric fashion. The guideline for such a generalization is
supersymmetry. Below we follow another way and generalize the
Lagrangian (\ref{estrla}) from a purely geometric point of view.
Namely the introduction of metric and torsion for a string surface
attributes new bosonic degrees of freedom describing the internal
structure of the string. The torsion degree of freedom is
dynamical. The choice of the geometric Lagrangian seems to be
unique in the framework of Riemann--Cartan geometry. There we have
looked over all invariants which yield the second order equations
of motion. Note that there is no geometric Lagrangian for a metric
alone without torsion which gives rise to the second order
equations of motion. The resulting theory has invariance (i) and
(ii) but Weyl invariance is broken already at the classical level.

It was shown \cite{Polyak81A} that the Weyl invariance of the model
(\ref{estrla}) is broken in the quantum domain. As a result, the conformal
piece of the metric participates in a quantum dynamics and the theory is
reduced to the quantum theory of the Liouville equation. The absence of
a simple vacuum solution of the Liouville equation presents difficulties
in the quantization of the theory. Attempts to construct a satisfactory
bosonic string theory have led us to the investigation of the following
possibility. The Lagrangian (\ref{estrla}) may be considered as describing
the two-dimensional gravity field $g_{\al\bt}$ interacting with the scalar
fields $X^\mu$. From this point of view the dynamical equation of motion
for the metric $g_{\al\bt}$ is needed.

Adding to the Lagrangian $L_0$ the Hilbert--Einstein Lagrangian $\sqrt gR$
does nor give rise to the dynamical equation for the metric because the
latter Lagrangian in two dimensions equals to a total divergence. But
if one admits a torsion to be nontrivial and dynamical then there exists
the possibility of obtaining a dynamical equation of motion for the metric
or for the corresponding zweibein. In conformal gauge this theory is reduced
to the theory of interacting scalar and vector fields and has a simple
vacuum solution.
\vskip5mm

2. The geometry of the string is described by the zweibein $e_\al{}^a$
$(a=0,1)$, $g_{\al\bt}=e_\al{}^a e_{\bt a}$ and the Lorentz connection
$\om_\al{}^{ab}=-\om_\al{}^{ba}$. Curvature and torsion have the following
form
\begin{equation*}
\begin{split}
  R_{\al\bt}{}^{ab}&=\pl_\al\om_\bt{}^{ab}-\om_\al{}^{ac}\om_{\bt c}{}^b
  -(\al\leftrightarrow\bt),
\\
  T_{\al\bt}{}^a&=\pl_\al e_\bt{}^a-\om_\al{}^{ab}e_{\bt b}
  -(\al\leftrightarrow\bt),
\\
  R_{abcd}&=R_{\al\bt cd}e^\al{}_ae^\bt{}_b,~~~~
  T_{abc}=T_{\al\bt c}e^\al{}_ae^\bt{}_b.
\end{split}
\end{equation*}

Let us consider the most general reparametrization-invariant and
parity-conserving Lagrangian quadratic in curvature and torsion
\begin{equation}                                        \label{elarts}
  L_1=\sqrt g\left(\frac14\mu^2R_{abcd}^2+\frac14\g^2T_{abc}^2+\lm\right),
\end{equation}
which yields the second-order differential equations. There are no
parity-violating terms in two dimensions giving rise to equations of
motion of no more then second order. Thus the Lagrangian (\ref{elarts})
represents the most general reparametrization-invariant Lagrangian which
yields the second-order equations of motion for the zweibein and the
Lorentz connection. The curvature squared term includes the kinetic one
for the Lorentz connection and the torsion squared term includes the
kinetic term for the zweibein and the mass for the Lorentz connection.
Note that only three parameters $\mu$, $\g$ and $\lm$ survive in two
dimensions, in contrast to the ten parameters \cite{SezNie80} of a
four-dimensional Lagrangian.

The string action
\begin{equation}                                        \label{estrac}
  S=\int d^2\zeta L,~~~~L=L_0+L_1
\end{equation}
yields the following equations of motion:
\begin{subequations}
\begin{align}                                           \label{eqmoxv}
  g^{\al\bt}\nb_\al\nb_\bt X^\mu&=0,
\\                                                      \nonumber
  -\g^2\nb_\bt T^{\bt\al}{}_a
  +\g^2\left(\frac14 T_{bcd}^2e^\al{}_a-T^{\al bc}T_{abc}\right)
  +2\mu^2\left(\frac14F_{bc}^2e^\al{}_a-F^{\al b}F_{ab}\right)+\lm e^\al{}_a&
\\                                                      \label{eqmzwf}
  +\frac12g^{\bt\g}pl_\bt X^\mu\pl_\g X_\mu e^\al{}_a
  -\pl^\al X^\mu\pl_\bt X_\mu e^\bt{}_a&=0,
\\                                                      \label{eqmolo}
  2\mu^2\nb_\bt F^{\bt\al}-\g^2 T^{\al ab}\ve_{ab}&=0.
\end{align}
\end{subequations}
Here $\nb_\al$ means a covariant derivative with Lorentz connection
$\om_\al{}^{ab}$ when it acts on the tensors with Latin indices, and with
the metrical connection without torsion $\G_{\al\bt}{}^\g$
(Christoffel's symbols) when it acts on the tensors with Greek indices
$\al$, $\bt$. We have parametrized the Lorentz connection by a vector
field $\om_\al{}^{ab}=A_\al\ve^{ab}$ where $\ve^{ab}=-\ve^{ba}$, and
$F_{\al\bt}=\pl_\al A_\bt-\pl_\bt A_\al$. We consider the string with
Euclidean metric. Note that we have the equation of motion (\ref{eqmzwf})
instead of the constraints in the usual approach. It leads to a
different quantization scheme. The corresponding Virasoro algebra will be
considered in a separate publication.

In conformally flat gauge $e_\al{}^a=e^\vf\dl_\al^a$ the Lagrangian
(\ref{elarts}) is reduced to the following expression up to a total
divergence:
\begin{equation}                                        \label{elacog}
  L_1=\frac12\mu^2e^{-2\vf}F_{\al\bt}^2+\frac12\g^2\left[(\pl_\al\vf)^2
  +\vf F_{\al\bt}\ve^{\al\bt}+A_\al^2\right]+\lm e^{2\vf}.
\end{equation}
It describes in flat Euclidean space the interaction of the scalar field
$\vf$ appearing from the zweibein and the vector field $A_\al$ appearing
from the Lorentz connection.

A special case of the Lagrangian (\ref{elacog}) is of particular interest.
Rescaling the fields $\vf\rightarrow\g\vf$, $A_\al\rightarrow\mu A_\al$
and taking the limit $\g/\mu\rightarrow0$ one has the following Lagrangian:
\begin{equation}                                        \label{escalt}
  L_1=\frac12e^{-2\vf/\g}F_{\al\bt}^2+\frac12(\pl_\al\vf)^2+\lm e^{2\vf/\g}.
\end{equation}
The remarkable property of this theory is its integrability. The Lagrangian
(\ref{escalt}) yields the following equations of motion:
\begin{subequations}                                    \label{eredem}
\begin{align}                                           \label{eqmvel}
  \pl_\al(e^{-2\vf/\g}F_{\al\bt})&=0,
\\                                                      \label{eqmphi}
  \triangle\vf=\pl_\al\pl_\al\vf&=
  \frac1\g\left(2\lm e^{2\vf/\g}-F_{\al\bt}^2e^{-2\vf/\g}\right).
\end{align}
\end{subequations}
Eq.~(\ref{eqmvel}) has the general solution
\begin{equation}                                        \label{esolfl}
  F_{\al\bt}=c\ve_{\al\bt} e^{2\vf/\g},
\end{equation}
with arbitrary constant $c$. Substitution of (\ref{esolfl}) into
(\ref{eqmphi}) yields the integrable Liouville equation
\begin{equation*}
  \triangle\vf=\frac2\g(\lm-c^2)e^{2\vf/\g}.
\end{equation*}
Thus the Lagrangian (\ref{escalt}) gives integrable equations of motion.
It is important that eqs.~(\ref{eredem}) have a simple vacuum solution
\begin{equation}                                        \label{evacso}
  \vf=\vf_0=\const,~~~~F_{\al\bt}=\sqrt\lm\ve_{\al\bt}e^{2\vf_0/\g},
\end{equation}
in contrast to the Liouville equation.

It is known that the Liouville equation closely relates to the
theory of surfaces with constant curvature. The Lagrangians
(\ref{elacog}) and (\ref{escalt}) seem to be related by analogy to the
geometry of surfaces with torsion. The Lagrangian (\ref{elarts})
is of interest also as a two-dimensional theory of gravity (about
lower dimensional theories of gravity, see ref.\cite{Jackiw84}).
\vskip5mm

3. Let us proceed to the quantization of the string model described by
the action (\ref{estrac}). The partition function has the following form
\begin{equation*}
  Z=\int\mathcal{D}(X^\mu)\mathcal{D}(e_\al{}^a)\mathcal{D}(\om_\al{}^{ab})
  e^{-S}.
\end{equation*}
We consider here only closed strings with the topology of a sphere.
The measure $\mathcal{D}(e_\al{}^a)$ takes the following form \cite{Polyak81A}:
\begin{equation*}
  \mathcal{D}(e_\al{}^a)=\mathcal{D}\vf\mathcal{D}\e^\al\mathcal{D}l
  \exp\left(-\frac{26}{12\pi}\int d^2\zeta
  \left[\frac12(\pl_\al\vf)^2+\lm_1e^{2\vf}\right]\right).
\end{equation*}
Integration over $X^\mu$ for a sphere gives \cite{Polyak81A}
\begin{align*}
  \int\mathcal{D}X^\mu\exp&\left(-\frac12\int d^2\zeta\sqrt g g^{\al\bt}
  \pl_\al X^\mu\pl_\bt X_\mu\right)=\exp(-\mathcal{F}),
\\
  \mathcal{F}&
  =-\frac D{12\pi}\int d^2\zeta\left[\frac12(\pl_\al\vf)^2+\lm_2e^{2\vf}\right].
\end{align*}
The above $\lm_1$ and $\lm_2$ are arbitrary constants. The measure
$\mathcal{D}(\om_\al{}^{ab})=\mathcal{D}(A_\al)$. Thus the partition function
takes the form
\begin{equation*}
  Z=\int\mathcal{D}\vf\mathcal{D}(A_\al)\exp\left(-\int d^2\zeta L_\mathrm{eff}\right),
\end{equation*}
where
\begin{equation*}
\begin{split}
  L_\mathrm{eff}&=\frac12\mu^2e^{-2\vf}F_{\al\bt}^2+\frac12\kappa^2(\pl_\al\vf)^2
  +\frac12\g^2\vf F_{\al\bt}\ve^{\al\bt}+\frac12\g^2A_\al^2+\tilde\lm e^{2\vf},
\\
  \kappa^2&=\g^2+\frac1{12\pi}(26-D),~~~~\tilde\lm=\lm+\lm_1-\lm_2.
\end{split}
\end{equation*}

The theory takes a particular simple form after rescaling the fields
$\vf\rightarrow\kappa\vf$, $A_\al\rightarrow\mu A_\al$ and taking the limit
$\g/\mu\rightarrow0$. Then the theory results in an integrable one with
an effective Lagrangian
\begin{equation}                                        \label{efflaq}
  L_\mathrm{eff}=\frac12e^{-2\vf/\kappa}F_{\al\bt}^2+\frac12(\pl_\al\vf)^2
  +\tilde\lm e^{2\vf/\kappa},
\end{equation}
as it was discussed earlier. This Lagrangian may be quantized in powers
of $1/\kappa$ around vacuum (\ref{evacso}) $\vf=\vf_0$,
$F_{\al\bt}=\sqrt{\tilde\lm}\ve_{\al\bt}e^{2\vf_0/\kappa}$ in any dimension
$D$ because the inequality
\begin{equation*}
  \kappa^2=\g^2+\frac1{12\pi}(26-D)>0
\end{equation*}
may be satisfied by an appropriate choice of $\g$.

So there exists an interesting string model which is described by the
action (\ref{estrac}) and seems to be very promising. Further details
of the present model, in particular the spectrum of the theory, will
be discussed in a separate publication.
\vskip5mm

One of the authors (I.V.) would like to thank I.~Ya.~Aref'eva and S.~Pacheva
for useful discussions. The author thanks the Leipzig University, where this
work was completed, for its hospitality.

\end{document}